\documentclass[12pt]{article} 
\def\be{\begin{equation}} \def\ee{\end{equation}}
\def\bi{\begin{itemize}} \def\ei{\end{itemize}}
\def\bea{\begin{eqnarray}} \def\eea{\end{eqnarray}} \def\ba{\begin{array}}
\def\ea{\end{array}} \def\ben{\begin{enumerate}} \def\een{\end{enumerate}}

\newcommand{\eqn}[1]{(\ref{#1})}

\newcommand{\prl}[3]{Phys. Rev. Lett. {\bf#1} ({#2}) {#3}}

\newcommand{\hepth}[1]{{\tt arXiv:{#1}[hep-th]}}

\def\br{\nonumber\\}

\def\Tr{{\rm Tr}}

\begin{document}
{}~
\hfill \vbox{
\hbox{\today}}\break

\vskip 3.5cm
\centerline{\Large \bf
 Entanglement entropy at higher orders
}
\centerline{\Large \bf
 for the states of $a=3$ $\theta=1$ Lifshitz theory
}

\vspace*{1cm}

\centerline{\sc Rohit Mishra and  Harvendra Singh
}

\vspace*{.5cm}
\centerline{ \it  Theory Division, Saha Institute of Nuclear Physics} 
\centerline{ \it  1/AF Bidhannagar, Kolkata 700064, India}
\vspace*{.25cm}
\centerline{ \it  Homi Bhabha National Institute, Anushakti Nagar, 
Mumbai 400094, India}
\vspace*{.25cm}

\vspace*{1.5cm}

\centerline{\bf Abstract} \bigskip
We evaluate the entanglement entropy of strips for 
boosted D3-black-branes compactified along the lightcone coordinate. 
The bulk theory describes 3-dimensional $a=3$ $\theta=1$ Lifshitz theory 
on the boundary. The area of small strips 
is evaluated perturbatively up to second order, where
the leading term has a logarithmic dependence on strip width $l$,   
whereas  entropy of the excitations is found to be proportional to $l^4$. 
The entanglement temperature falls off 
as ${1/ l^3}$ on expected lines. The size of the subsystem has to be bigger
than typical `Lifshitz  scale' in the theory.
At second order, the redefinition of temperature (or strip width) 
is  required so as to meaningfully describe the entropy corrections 
in the form of a first law of entanglement thermodynamics.       

\vfill 
\eject

\baselineskip=16.2pt


\section{Introduction}

The AdS/CFT correspondence \cite{malda} has remained a central
 idea for  holographic studies in string theory. The holography
 relates conformal field theory
living on the boundary of   anti-de Sitter
spacetime with  the gravity theory within the bulk. 
Along  these lines finding the entanglement entropy of strongly coupled
quantum systems at
criticality has also been a focus of several studies  \cite{RT,ogawa}.
In these calculations 
 entanglement entropy can  be obtained \cite{RT,Hubeny:2007xt}
 by estimating the  area of  codimension two surfaces embedded 
inside the  bulk  geometry. The boundary of such extremal surfaces coincides
with the boundary of the subsystem in the CFT.
Recently it has been observed 
that the  excitations  in the CFT  
follow  entanglement   laws 
similar to the black hole thermodynamic laws \cite{JT, alisha, ms2015};
 see also  \cite{Pang:2013lpa}, 
\cite{Wong:2013gua}, \cite{Park:2015afa}, \cite{Ghosh:2017ygi}.
It is  understood now that the entanglement entropy ($S_E$) and 
the energy of small excitations (${ E}$) in  AdS spacetime obey
a definite relation
$$\bigtriangleup { E}= 
T_E \bigtriangleup S_E  +
{\cal V} \bigtriangleup {\cal P}  +
\mu_E \bigtriangleup N   
$$
This equation is described as the first law of entanglement 
thermodynamics. The charge 
contributions can simply  arise for a boosted black-brane vacua \cite{ms2015},
where the charged excitations could be either  Kaluza-Klein (KK)
momentum modes along a compactified brane direction or  
the dual winding modes of a string.

The  backgrounds of our interest here  are  
the nonrelativistic Lifshitz  spacetimes. We would like to holographically
study these solutions and check if similar entanglement law 
could be written for them.
Typically a Lifshitz like geometry \cite{kachru} has a line element
$$ds^2=-{dt^2\over z^{2a}}+{dx_1^2+\cdots+dx_D^2 \over z^2}+{dz^2\over z^2}
$$
as subspace where time and space  scale asymmetrically  
($z\to\lambda z,~t\to\lambda^a t,~ x_i\to\lambda x_i$). The
 Lorentz symmetry is explicitly 
broken. The parameter $a$ is called the dynamical exponent of time and $D$
is total spatial dimensions.
Whereas an hyperscaling violating Lifshitz (hvLif) geometry 
 $$ds^2=z^{\theta}
\left(-{dt^2\over z^{2a}}+{dx_1^2+\cdots+dx_D^2 \over z^2}+{dz^2\over z^2}
\right)$$
has an overall conformal factor \cite{ogawa}. So  an overall scaling
$g_{\mu\nu}\to \lambda^\theta g_{\mu\nu}$ of the metric is involved.
As an unique example in ten dimensions, 
especially $Lif^{a=2}_4\times S^1\times S^5$ vacua
are  recently constructed in \cite{Singh:2017wei}, 
as  solutions of 10-dimensional massive type IIA supergravity
 theory \cite{roma}. These are understood to describe strongly coupled
 Lifshitz $a=2$ theory
in three spacetime dimensions at the fixed point.
In these bulk solutions
`massive' strings are tied up with D2-D8 parallel brane system that  
 exhibit scaling symmetry. 
 These vacua can be related via massive/generalised  T-duality \cite{berg}
 to  D3-D7 axion `flux' vacua  \cite{kbala}.
In the ordinary type IIA/B string theory and  M-theory, 
the boosted black brane solutions
compactified along a lightcone direction,
can also give rise to Lifshitz  solutions  \cite{hs2010, hs2012}. 
The latter class of brane solutions all  have  conformal 
scaling (or hyperscaling)
properties as listed in \cite{hs2012},
see \cite{knara} for discussion of $\theta=1$ case. 
There are other  instances also
in gauged supergravities where Lifshitz vacua can  be obtained, see
for example \cite{Taylor:2015glc}.

In this work we shall only study  boosted black D3-brane system in lightcone 
coordinates,  with one lightcone coordinate compactified 
on a circle \cite{maldatachi}. These 
compactified bulk solutions   describe a thermal state of 3-dimensional
 $Lif^{a=3}$ theory with hyperscaling parameter $\theta=1$. The corresponding
ground state  (at zero temperature) is  described by a ten-dimensional
solutions  discussed in \cite{hs2010}. These zero temperature
solutions were obtained by performing a double scaling limit in which horizon
size is taken to vanishing value associated with an infinite boost.
 The fact that there is 
hyperscaling violation in these Lifshitz solutions
(when explicit lightcone 
compactification is performed) was pointed out 
subsequently in the work \cite{knara}. Actually
these  hvLif theories fall in 
a special category where $\theta=D-1$ ($D=2$ is number of spatial dimensions). 
Precisely for  these
hvLif states the entanglement entropy has logarithmic violation 
\cite{ogawa, sachdev1111, dong1201}. 
We shall be encountering some of
these results in our work as we progress.
 Both the solutions (finite temperature and the zero temperature one) 
allow us to embed codimension-2 strip like 
surfaces (at constant lightcone time)  inside the bulk.  
We evaluate the area of small strips using perturbative
method up to second order, by using the procedure
 introduced in  \cite{ms2015}. We find that for small strips 
(but bigger than some critical size) 
the leading term in the entanglement entropy has indeed 
logarithmic dependence on strip width 
$l$. Whereas the subleading term which accounts for 
the entropy of excitations  goes as $l^4$. The entanglement
temperature falls off as ${1\over l^3}$. These results are  on  expected 
lines. Quite importantly, the (KK) charge contribution in the first law 
is present at the
first order itself, unlike in the relativistic cases studied in 
\cite{ms2015} where
the contibution of the charges  appears only at the second order. 
At the second order, once again we  find that the first law relation requires 
the entanglement temperature 
(and strip width)  to be suitably corrected or renormalised. 
    
The paper is presented as follows. In section-2 we write down the 
$Lif^{a=3,\theta=1}$ hyperscaling solutions of our interest both 
with black hole excitations
and the zero temperature counterpart. In section-3 we  evaluate the
entanglement entropy at first order and present the form of first law.
In section-4 we obtain  second order corrections and rewrite the 
new form of first law and determine the corrections to 
associated  thermodynamic quantities. We find
that the strip width (so also  subsystem volume) has to be 
renormalised and redefined. 
The final summary is presented in section-5.

\section{Entanglement entropy for $a=3$ $\theta=1$ Lifshitz system}
It has been known that 
the boosted $AdS_5\times S^5$ black hole background compactified along
a light-cone coordinate can describe  excitations of $hvLif^{a=3}$ system
\cite{hs2010,hs2012}. These black hole
solutions were first explored for their
 non-relativistic properties in \cite{maldatachi}. 
These type IIB string vacua can be written as 
\bea\label{bst1}
&&ds^2=L^2\left( -{ z_l^4 f \over z^6} (dx^+)^2+
{z^2\over 4 z_l^4} 
 (dx^- - \omega)^2
+{dx_1^2+dx_2^2\over z^2}+{dz^2 \over f z^2}\right) 
 + L^2 d\Omega_5^2,
\eea
 supported by  constant dilaton 
and a self-dual 5-form field strength. 
The function $f$ is
\be
 f(z)=1-{z^4\over z_0^4}, ~~~~~
\ee
where  $z=z_0$ is  the black hole  horizon. 
It will  be  assumed that $x^-$ is compactified on a circle 
of radius $r^-$. The fiber 1-form is given by
\be
\omega={z_l^4\over  z^4}(2-{z^4\over z_0^4}) dx^+ 
\ee
The radius of curvature $L$
 is taken very large  in  string units ($\alpha'= 1$) so that the stringy excitations
are suppressed. 
The parameter $z_l$ is an intermediate (free) UV 
scale, rather we shall suitably call it as `Lifshitz scale' in the theory. 
We  take a wide parameter range such that $z_0\gg z_l$. 
This is so because we wish to study small excitations only. 
(Also let us  note  that at any stage the Lifshitz scale $z_l$  can be  
related to $z_0$ through the boost of  lightcone coordinates, 
i.e. one can write $z_l^2 =z_0^2/\lambda$, with
 $\lambda \ge 1$ being  the lightcone boost parameter.) 
 
Further we shall take the $a=3$ $\theta=1$ Lifshitz solutions \cite{hs2010} 
as the ground state. Let us explain it here. 
Recalling \cite{hs2010}, one can take simultaneous  double limits $
\lambda\to \infty, ~
z_0 \to\infty$, while keeping the ratio 
${\lambda\over 
z_0^2}={1\over z_l^2}$ (say) fixed. These limits 
take us to (hyperscaling) Lifshitz $a=3$ vacua, namely
\bea\label{lifa3}
ds^2_{Lif}
&=&{L^2}\left( -{z_l^4  (dx^+)2\over z^6}+
{z^2\over 4z_l^4}  (dx^- - {2 z_l^4\over z^4}dx^+)^2
+{dx_1^2+dx_{2}^2+dz^2 \over z^2 } + d\Omega_5^2\right) \br
&\equiv&{L^2}\left( -{dx^+dx^-\over z^2}
+{z^2\over 4z_l^4}  (dx^-)^2
+{dx_1^2+dx_{2}^2+dz^2 \over z^2 } + d\Omega_5^2\right) 
\eea
This  zero temperature
background is characterized by  the scale $z_l$,
which also defines the  charge (number) density of the states of
this system at zero temperature \cite{hs2010}. It only suggests that 
 $a=3$ hyperscaling Lifshitz  ground state  system 
exists for any given  $z_l$. The 
$z_l$ is treated as Lifshitz (intermediate) 
scale in the black hole solution 
\eqn{bst1}, when we switch on the temperature. That is we are interested
in the  excitations around the hyperscaling Lifshitz vacua \eqn{lifa3}. 

The entanglement entropy has also been studied for these BH systems
in the work \cite{narayan}. It was pointed out there that 
due to the boost there is an asymmetry in the entanglement 
along various directions of the boundary theory. This asymmetry 
should show up in the entanglement
entropy calculations and the first law as well. Up to  first order 
in  perturbative expansion (for small subsytem) 
it has been explicitly shown that it  indeed is the case 
\cite{Mishra:2016yor}.
The observation was that the entanglement asymmetry is entirely
due to pressure asymmetry in the theory. But in the current paper
 we shall be studying only 
those  strip sytems which lie in the transverse to the boost direction,
it is because the solutions are compactified along the boosted lightcone
coordinate, so that we can view it as a hyperscaling Lifshitz theory. 
The entanglement 
entropy along the boost direction can also be studied
at higher orders, but this will require some careful considerations, because
the constant time slices will not be existing. Technically one has to 
resort to a covariant slice
analysis \cite{Hubeny:2007xt}. 
 Up to first order the calculations are all easy and there are 
no such hurdles. But beyond first order we have to only consider 
covariant approach such as 
\cite{Hubeny:2007xt}. 
We leave it for a separate investigation.

\subsection{Small strip systems}

The entanglement  entropy for a subsystem on the 
boundary of the  background  \eqn{bst1} can be studied
by using Ryu-Takayanagi proposal \cite{RT}. Here $x_1$ and $x_2$
 are two flat directions along the brane, while spatial lightcone coordinate 
$x^-$ is compactified. 
We choose a strip  along $x_1$ direction with
 an interval  $-l/2 \le x_1 \le l/2$.
We wish to embed  co-dimension two strip  
(a constant $x^+$ surface) inside  the bulk  geometry. 
The two straight boundaries of the
extremal strip  surface   coincide with the two ends 
of the interval $\bigtriangleup x_1=l$.   
The  size of  other  coordinates are taken as; 
 $ x^-\simeq x^-  + 2 \pi r^-$, 
$0\le x_2\le l_2$, where $l_2$ 
is taken very large, $ l_2\gg l$. 

Following Ryu-Takayanagi prescription 
the entanglement entropy of a strip subsystem  is given 
in terms of the geometrical area of co-dimension two surface
(with light-cone time  $x^+$ taken constant everywhere on the
surface). We thus get 
\bea\label
{schkl1saa}
 S_E &&\equiv {{\cal A}_{Strip}\over 4G_5} \br
&&=
 {  L^3\pi r_{-}l_2 \over 
2G_5 z_l^2}
\int^{z_\ast}_{\epsilon}dz  { 1\over z  \sqrt{f}} \sqrt{1 +f(\partial_z x^1)^2}
\eea  
where  $G_{5}$ is $5$-dimensional Newton's constant. 
Here $\epsilon \sim 0$,  is the cut-off scale in UV.
(We need to pay special attention to  $z=z_l$ scale in the theory.
The bulk geometry \eqn{bst1} 
is not well defined  beyond $z=z_l$, as the size of the $x^-$ circle
becomes sub-stringy in $z<z_l$ near boundary region. 
A way to overcome this problem is that beyond 
$z=z_l$ one can switch over to  T-dual type-IIA  background, where 
the circle size instead will increase.
Doing this however does not  affect the entropy functional given
in \eqn{schkl1saa}.
Hence so far as the area functional is concerned it appears immune to 
$z=z_l$. Nevertheless $z_l$ is an important scale in the Lifshitz theory and
 we can  add suitable counter terms as we shall discuss next.). 
The $z_\ast$ is the turning point
of the  strip.  Next
the  area functional  is extremized 
through the equation of motion 
\be\label{klop}
{dx_1 \over dz}=  {z\over z_\ast \sqrt{f}} 
 {1\over\sqrt{1- ({z\over z_\ast})^2}} 
\ee
It implies that the boundary value $x^1(0)=l/2$ 
is given by the following integral
relation 
\be\label{klop1}
{l \over 2}=  \int_{0}^{z_\ast} dz
{z\over z_\ast} {1 \over \sqrt{f} 
 \sqrt{1- ({z\over z_\ast})^2}} 
\ee
which relates the width $l$ with the turning point $ z_\ast$.   
The turning-point of the strip 
lies at the mid-point  $x^1(z_\ast)=0$ of the boundary interval 
due to symmetry. The  entropy
for extremal  strip system can now be described as 
\be\label{kl1kv5}
S_E= {  L^{3} \pi r^-l_2\over 2 G_{5}z_l^2}
\int^{z_\ast}_{\epsilon}{dz \over  z}{1 
\over\sqrt{f}}{1\over \sqrt{1-({z\over z_\ast})^{2}}} 
\ee  
The Lifshitz scale
$z_l$ is an important fixed parameter in these vacua, but it 
only appears as a constant multiplier outside the integrand.

\subsection{A hierarchy of  scales and perturbative expansion}
When  strip width $l$ is  small 
the turning point  generically lies in the proximity of
 asymptotic  region. Therefore one can safely  assume
 $z_\ast \ll z_0$. However, our  main
focus here will be  on those subsystems (or critical surfaces) for which  
following hierarchy of scales is obeyed
\be\label{hier1}
z_l< z_\ast \ll z_0 \ .
\ee
This will specially require us to take $z_l$ (UV) and $z_0$ (IR)  to be
widely separated scales. This $[z_0,z_l]$ interval is known as 
 the `Lifshitz window' region in \cite{hs2012,hs2011}. A
large Lifshitz window is desirable here for   perturbative expansion 
to work out properly, as
we are seeking to evaluate the entanglement entropy 
\eqn{kl1kv5} by expanding it around a zero temperature 
 vacua \eqn{lifa3} (i.e. 
treating $a=3$ $\theta=1$ Lifshitz vacua \cite{hs2010} as the  ground state).
 Under these conditions we can estimate  area entropy perturbatively by 
expanding the integrand around its central value. 

We first proceed to obtain
the perturbative expansion of the $l$-integral \eqn{klop1} up to first order,
assuming ${ z_\ast^4\over z_0^4}\ll 1$, 
\bea
&&l=2 z_\ast \int_0^1 d\xi {\xi \over\sqrt{1-\xi^2}} 
(1+ { z_\ast^4\over 2 z_0^4} \xi^4 
+~ \cdots)\br
&&= 2z_\ast b_0 + {z_\ast^5  \over  z_0^4}{b_1} 
+~ \cdots
\eea
where for simplicity we  introduced
$\xi\equiv {z\over z_\ast}$ and $ R\equiv  (1-\xi^2)$
.\footnote{
The value of  expansion coefficients $b_0,b_1,b_2$  
can be  evaluated,
$ b_0=\int_0^{1} d\xi {\xi  \over\sqrt{ R}}
={1\over 2}B(1,1/2)
=1 ,~~~
 b_1=\int_0^{1} d\xi {\xi^5 \over \sqrt{ R}}
={1\over 2}B(3,1/2)
={8\over 15}, ~~~
 b_2=\int_0^{1} d\xi {\xi^9 \over \sqrt{R}}
={1\over 2}B(5,1/2),$
where $B(m,n)={\Gamma(m)\Gamma(n)\over\Gamma(m+n)}$ are the Beta-functions.}
 The ellipses stand for second and higher order terms which we neglect
in this section.
By inverting  the above  series we get a turning point expansion
 \bea \label{turnp1}
z_\ast= \bar z_\ast(  1-{\bar z_\ast^4\over  z_0^4}{b_1\over2b_0} )+\cdots
\eea
where $\bar z_\ast\equiv {l \over 2b_0}$ 
is the turning point for pure hvLif ground state \eqn{lifa3} (i.e.
in the absence of excitations or black holes). This relationship 
is an important first step before we proceed to the area calculation.

Next we consider the area  of the strip \eqn{kl1kv5}. We
 evaluate the  integral quantity (which is independent of $z_l$)
\be\label{kl1kv}
A\equiv  
{2\over z_l^2}
\int^{z_\ast}_{\epsilon}{dz \over  z}{1 
\over\sqrt{f} \sqrt{1-({z\over z_\ast})^{2}}} 
\ee  
by expanding the integrand perturbatively as 
\bea\label{kl1kvn}
A\equiv  {2\over z_l^2}(
  \int^1_{\epsilon/z_\ast}
{d\xi\over\xi}{1\over\sqrt{ R}}
+  {z_\ast^4\over 2z_0^4}
 \int_{\epsilon/z_\ast}^1 d\xi{\xi^3\over \sqrt{R}} 
+\cdots).
\eea
The contribution of first term
is singular when $\epsilon\to 0$ (near the boundary). Also as mentioned before,
going beyond $z=z_l$, the $x^-$ circle in  \eqn{bst1}
becomes sub-stringy, so
near boundary region $z<z_l$ needs to be carefully considered. 
We thus note that, in the corresponding
dual geometry  the size of T-dual $x^-$ circle
  will anyway expand for $z<z_l$. While 
 the functional form of integral in eq.\eqn{kl1kv} 
remains unchanged under this duality. 
Thus there appears to be  no pathological problem in the near boundary region
$0<z< z_l$. Nevertheless,
to be on the safe side 
we  subtract the following contribution (as a counter term) 
\be
A_{CT}=
{2\over z_l^2}\int_\epsilon^{z_l}  {dz\over z}=
{2\over z_l^2}\ln{z_l\over\epsilon}
\ee
from the area integral $A$ given above. This precisely amounts to subtracting
the contribution of two  disconnected (no turning point) strips
 hanging between $z=z_l$ and the
 $z=\epsilon$ inside the hvLif  geometry \eqn{lifa3}. 
Note that $A_{CT}$  has no dependence on $z_0$, which is a parameter  
controlling the  excitations. So it is totally a harmless subtraction 
from  point of view of the  excitations 
(our goal is to know the entropy
 of the excitations and it  will not be affected). 
So we extract the  finite area contribution as
\bea \label{pol2}
A_{finite}=A-A_{CT}&&={2\over z_l^2}(
  \int^1_{\epsilon/z_\ast}
{d\xi\over\xi}{1\over\sqrt{ R}}
+  {z_\ast^4\over 2z_0^4}
 \int_{\epsilon/z_\ast}^1 d\xi{\xi^3\over \sqrt{R}} 
+\cdots) -{2\over z_l^2}
\int_{\epsilon/z_\ast}^{z_l/z_\ast}  {d\xi\over \xi} \br
&&= {2\over z_l^2} \ln {2 z_\ast\over z_l}
 +{1\over z_l^2}{z_\ast^4\over z_0^4}
 \int_0^1 d\xi{\xi^3\over \sqrt{R}}+\cdots
\eea
 and the limit $\epsilon\to 0$ 
is understood to have been
implemented in the second equality. In the
next step by substituting the expansion of  
$z_\ast$ in the eq.\eqn{pol2}, we get  up to
first order
\bea
A_{finite}&
=& {2\over z_l^2} \ln {2\bar z_\ast\over z_l} -{2\over z_l^2}
{\bar z_\ast^4\over  z_0^4}
{b_1\over 2} + {2\over z_l^2} {\bar z_\ast^4 \over  z_0^4} 
{a_1\over 2}\br
&=& 
 {2\over z_l^2} \ln {l\over z_l}
+{1\over z_l^2}
{\bar z_\ast^4\over  z_0^4} (a_1-{b_1}) \br
&=& A_0
+{1\over z_l^2}
{\bar z_\ast^4\over  z_0^4} {a_1\over 5} 
\eea
where $a_1, b_1, \cdots$   are finite coefficients.\footnote{The 
expansion coefficients are
$ a_1=\int_0^{1} d\xi {\xi^3 \over   \sqrt{1-\xi^{2}}}
={1\over2}B(2,1/2)={2\over3}, ~~ 
a_2=\int_0^{1} d\xi {\xi^7 \over   \sqrt{1-\xi^{2}}}=
{1\over2}B(4,1/2)={6\over 7} b_1 
$.}
The leading finite  term is simply given by
\bea
A_0={1\over z_l^2}
 \ln {l^2\over z_l^2} 
\eea
Thus the  entanglement entropy for   strip can be written as
\bea\label{hj4}
 S_E =  S_{(0)}+
 {L^{3} 
\pi r_{-}l_2 \over 4 G_{5}} {a_1\over5  z_l^2} {\bar z_\ast^4\over  z_0^4}
\eea
where the leading term is
\be
S_{(0)}= 
{L^{3} \pi r_{-}l_2 \over 4 G_{5}} 
{1\over  z_l^2} \ln {l^2\over z_l^2} 
\ee
It is  clear that $S_{(0)}>0$  only when $l >z_l$ and that is why the 
hierarchy of the scales \eqn{hier1} was adopted. It  
also does not look
like  an  $AdS_5$ ground state entropy which instead goes as 
$-{1\over l^2}$ \cite{RT}. Therefore the logarithmic dependence on  $l$ 
 ought to be recognized as a contribution of
 ${a=3},\theta=1$ Lifshitz ground state \cite{ogawa,sachdev1111}. 
This  also  happens because we have  chosen 
to study $x^+=constant$  strip subsystems.  
Had we chosen to evaluate entanglement entropy for usual 
$x^0=constant$ (fixed Lorentzian time) strips, we  instead  would get 
the leading 
contribution precisely  that for 
 $AdS_5$;    see \cite{ms2015} for  a second order  
perturbative calculation in relativistic theory). 

The leading logarithmic term 
 depends  on $z_l$ (UV scale)  and  the  width $l (> z_l)$, 
and not on $z_0$ (the scale describing the excitations).
But both of these quantities are fixed for a given subsystem. Thus
  $S_{(0)}$  is essentially a fixed quantity and it cannot be
viewed as part of the excitations.  
Subtracting the leading  term 
 leaves us with the net {\it vacuum-subtracted } 
  entropy  of the  excitations around Lifshitz theory as
\bea\label{hj4j}
\bigtriangleup S_E^{(1)} =  
 {L^{3} \pi r_{-}l_2 \over4 G_{5}} {a_1\over5  z_l^2} 
{ l^4\over  (2z_0)^4}+ {\rm higher~ order ~corrections}
\eea
This result is true up to first order in the ratio 
${ \bar z_\ast^4\over z_0^4}$.
At higher order there will be further corrections on the right hand side
to add.
It can be immediately observed that the entanglement entropy
of excitations  is proportional
to $l^4$ and  depends on $z_0$ also, the parameter describing 
excitations. 
In contrast, for  Lorentz covariant $AdS_{d+1}$ ground state 
the  entropy 
of excitations rather increases quadratically  as $l^2$ \cite{JT}.  

\section{ The  entanglement first law }

The boundary theory is a $3$-dimensional Lifshitz theory, since
 the lightcone direction, namely $x^-$, is compactified.
The excitation energy and the  pressure 
 can be obtained by expanding the  geometry \eqn{bst1} in
  Fefferman-Graham  coordinates  near the  boundary \cite{fg}. 
The energy density of the excitations is given by \cite{maldatachi}
\bea\label{hj5}
&&  {\cal E}
=  {  L^3 r_{-}  \over 16 G_{5}} {1\over 
z_0^4},
\eea
whereas the  charge (number) density is 
\bea
&& 
  {\cal \rho}=
{N\over volume}=  {  L^3 r_{-}^2  \over 8 G_{5}} 
{1\over  z_l^4}
\eea
The charge density in the Lifshitz theory at zero temperature
 is usually  
  very large  whereas other quantities can be vanishingly
small \cite{hs2010}. It is  obvious here too as
$\rho \propto  {1\over  z_l^4}$ and given our 
 hierarchy of the scales $z_l< z_\ast \ll z_0$.
The 'entanglement' chemical potential,  obtained by measuring the value 
of KK field $\omega_+$ at the turning point, is 
\bea
\mu_E
&=& {1\over r_-}(2 {z_l^4\over z_\ast^4} -{z_l^4\over  z_0^4}) 
= {1\over r_-}( {2z_l^4\over \bar z_\ast^4} 
+({4b_1\over b_0}-1){z_l^4\over  z_0^4}) \br
&=&  \mu_E^{Lif}
+ {1\over r_-}({4b_1\over b_0}-1){z_l^4\over  z_0^4} 
\eea
where to obtain
 second equality the turning point expansion \eqn{turnp1} has been used.
The leading term $ \mu_E^{Lif}= 
{1\over r_-} {2z_l^4\over \bar z_\ast^4}$ 
is   chemical potential corresponding to 
the hvLif ground state \eqn{lifa3}. The 
subleading term is however of {\it
universal} nature, because it is independent of $l$. 
Thus the net change in  chemical potential due to excitations is
\bea
\bigtriangleup \mu_E^{(1)}=\mu_E- \mu_E^{Lif}
\simeq  {1\over r_-}({4b_1\over b_0}-1){z_l^4\over  z_0^4} 
\eea
It is remarkable that,
using the quantities defined so far,
from \eqn{hj4j} we  can construct the following  first law-like relation
\bea\label{tempo1}
\bigtriangleup 
S_E^{(1)} =  {1\over T_E}(
\bigtriangleup E +{1\over 2}  
N\bigtriangleup \mu_E^{(1)} )
\eea
where the net charge contained in the subsystem is simply 
$N=\rho l_2 l $ and energy of excitations
$\bigtriangleup 
E=l_2 l {\cal E}$. The entanglement temperature is given by
\be\label{temp1av}
T_E= {2^6 z_l^2\over  \pi}{1\over l^3} .
\ee
Importantly the temperature
 is inversely proportional to the cubic power of the strip width $l$. 
This  conveys the fact
 that the dynamical exponent of time for the Lifshitz theory is indeed
$three$, and it corroborates  with  early work  \cite{hs2010}. 

We add some remarks here.  In the first law \eqn{tempo1}
the charge and chemical potential contribution is 
present at the first order itself, unlike in the relativistic case
where no charge  appears at the first order. In the relativistic case
the charges appeared only at the second order
 in perturbation, see \cite{ms2015}. 
The  reason for this major difference may be the fact that
the charge  density is very high in the Lifshitz theory, i.e.
$$  {{\cal \rho}
   \over
{\cal \rho}_c}
={z_0^4\over  z_l^4}\gg 1 \ .$$
where ${\cal \rho}_c=
{L^3 r_{-}^2  \over 8 G_5}{1\over z_0^2}$ is some critical (reference)
charge density.

The von Neumann entanglement 
entropy $ S_E =-\Tr \sigma_A \ln \sigma_A$ 
of a quantum subsystem $A$ requires the knowledge of 
a reduced density matrix (obtained by tracing out the states over the 
complimentary system),
 \be
\sigma_A={e^{-H_A}\over Z} \ee
where partition function $Z= \Tr_A e^{-H_A}$.
The $H_A$ is the reduced Hamiltonian describing  the
subsystem. In this approach, at first order 
we expect that the modular (entanglement) Hamiltonian $H_E$ 
of the subsystem
to be related as, \cite{casini, Wong:2013gua},   
\bea\label{tempo2}
\bigtriangleup S_E^{(1)} =  
{1\over T_E}(
\bigtriangleup E +{1\over 2}  N
\bigtriangleup \mu_E )=<\bigtriangleup H_E^{(1)}>  . 
\eea

\noindent{\it \large A variational form of first law:}\\
The small fluctuations  of bulk 
parameters $(z_0, z_l)$  determine the variations
of the thermodynamic quantities of boundary nonrelativistic  theory. 
In the present  $hvLif^{a=3}$ case 
 we are interested in the  study of the ensembles
 with fixed KK charges which can only be done  
 by keeping  $z_l$  fixed,  so we will only
 allow $z_0$ to have a spread.  
The small variation of  chemical potential becomes 
(at first order)
\be
\delta \mu_E
= (4b_1 -1){z_l^4\over r_-}\delta({1\over  z_0^4}) 
\ee
as given $b_0=1$.
One can  see that the  product 
$ {\cal \rho}.\delta \mu_E$ is of the same order as $\delta {\cal E}$
and thus it will eventually
 be related to it.
Hence the states of the system describe  a canonical ensemble and therefore
 knowing the fluctuations of a single  quantity, such as
 $\delta {\cal E}$,
is sufficient to describe the  state of the system. 
Under these restrictions (since $\delta S_{(0)}=0$ as $z_l$ is fixed) 
we find from eq. \eqn{hj4} the variational form of first law  is 
\bea \label{alis1b}
&& \delta  S_E  
= {1\over T_E} \delta{E'}
\eea
where new energy $E'\equiv E+{1\over 2}  \mu_E N$ has been defined so that
the entanglement temperature is the same as \eqn{temp1av}.
For a comparison with the black hole first law, we wish to recall 
the thermal first  law  \cite{maldatachi} 
for  boosted BH  background, which for fixed charge density ($z_l$=fixed),
gets reduced to
\bea \label{alis1c}
&& \delta  S_{th}  
= {1\over T_{th}} \delta{ E} 
\eea
 where 
\be
T_{th}={z_l^2\over \pi z_0^3}
\ee
is thermal temperature. 
It is worthwhile to note that not only the $z_l$ dependence in  entanglement
temperature is exactly the same as that in  thermal temperature but 
the dynamical exponent of time  also comes out as $3$.
Usually for smaller subsystems the 
entanglement temperature is  higher as compared to the thermal one
(if any). It is appropriate to compare the two in the present Lifshitz case.  
The ratio comes out to be
\bea
{T_{th}\over T_E}= 
{1\over 8} ({l \over  2z_0})^3   \ll 1
\eea
Since $l \ll 2z_0$, there will exist
 a big hierarchy in  two temperature scales where the degree is
determined by the dynamical exponent of time. 
Though the   ratio  remains independent 
of $z_l$ (or charge density of Lifshitz states), but 
it crucially  depends on the value of 
dynamical exponent, which obviously enhances this hierarchy. 

\section{ The entropy at second order}

We  wish to evaluate the area  of the lightcone strip
up to next higher order and evaluate corresponding entropy corrections. 
The higher order results provide us with better precision and improved estimate
of the entanglement entropy since exact analytical 
calculations cannot be done. First
the expansion of the turning point has to be obtained. We expand the
integrand in eq.\eqn{klop1} up to second order in ${z_\ast^4\over z_0^4}\ll1$,
which gives us the series
\bea
l &=&2 z_\ast \int_0^1 d\xi {\xi \over\sqrt{1-\xi^2}} 
(1+ { z_\ast^4\over 2 z_0^4} \xi^4 
+{3 z_\ast^8\over 8 z_0^8} \xi^8 
+~ higher~orders)\br
&&= 2z_\ast (1 +
{z_\ast^4  \over  z_0^4}{b_1\over 2} 
+{z_\ast^8  \over  z_0^8}{3b_2\over 8}) 
+~ \cdots
\eea
where the ellipses stand for third and  higher order terms.
By inverting the above expansion one can obtain 
 \bea \label{kj4}
z_\ast= \bar z_\ast \left( 1+
{\bar z_\ast^4\over  z_0^4}{b_1\over 2}
+{\bar z_\ast^8\over  z_0^8}
({3\over 8}{b_2}-{b_1^2})\right)^{-1}
\eea
where $\bar z_\ast=l/2$ is the turning point value 
for the ground state.
 The $A$ expansion up to second order 
is, keeping the   counter term same as
in the previous section, 
\bea \label{pol2o}
A-A_{CT}&=&{2\over z_l^2}(
  \int^1_{\epsilon/z_\ast}
{d\xi\over\xi}{1\over\sqrt{ R}}+
  {z_\ast^4\over 2z_0^4}
 \int_{\epsilon/z_\ast}^1 d\xi{\xi^3\over \sqrt{R}} +
  {3z_\ast^8\over 8z_0^8}
 \int_{\epsilon/z_\ast}^1 d\xi{\xi^7\over \sqrt{R}} 
+\cdots ) -{2\over z_l^2}\int_{\epsilon/z_\ast}^{z_l/z_\ast}{d\xi\over \xi}\br
&=& {2\over z_l^2} (\ln { 2z_\ast\over z_l}
 +{z_\ast^4\over 2z_0^4}
 \int_0^1 d\xi{\xi^3\over \sqrt{R}}
 +
  {3z_\ast^8\over8 z_0^8}
 \int_0^1 d\xi{\xi^7\over \sqrt{R}}
+\cdots)\br
&=&  {2\over z_l^2} \ln {2 z_\ast\over z_l} 
+ {1\over z_l^2} { z_\ast^4 \over  z_0^4} 
{a_1}
+ {3\over 4z_l^2} { z_\ast^8 \over  z_0^8} 
{a_2}
\eea
where $\epsilon\to 0$ limit has been implemented. The
coefficients $a_1, a_2$  are  defined earlier.
Substituting the  $z_\ast$-expansion  \eqn{kj4}  in the  $A$
expansion \eqn{pol2o}, we finally get
\bea
&&A_{finite}
=A_0+A_1+ A_2
\eea 
where $A_0$ and $A_1$ are the leading order and first order terms, respectively.
These are the same as obtained in the previous section. The new 
 contribution at second order is
\bea
A_2\equiv
{1\over 4 z_l^2} ({9}{b_1^2} -8a_1 {b_1}
 -{3} ({b_2}-a_2))
{\bar z_\ast^8\over  z_0^8}
\eea
With this the entanglement  entropy  calculated up to
 second order  becomes
\bea
 S_E& =&  
 S_{(0)}+  
 {L^{3} \pi r_{-}l_2 \over 4 G_{5}}(A_1+A_2).  \eea
So overall change up to second order is
\bea\label{hj4nb}
\bigtriangleup 
S_E^{(2)}
&=& S_E-S_{(0)}=  
 {L^{3} \pi r_{-}l_2 \over 4  G_{5}}\cdot  
{a_1 Q\over5 z_l^2 } \cdot
{l^4\over 2^4 z_0^4} 
\eea
where $Q$ factor  is  defined  as
\be
Q=(1- {26\over 105} { \bar z_\ast^4\over z_0^4}) <1
\ee
which involves  first order term only. It is always smaller than unity. 
This in fact fact implies that overall entanglement entropy after 
the inclusion of 
 second order corrections has indeed decreased,
\be
\bigtriangleup 
S_E^{(2)}<
\bigtriangleup 
S_E^{(1)}.
\ee
This is a  common observation in many CFTs including  the 
relativistic ones.
This calculation ends our perturbative results up to second order. In the next
step we would like to see if the second order corrections can be absorbed in
the redefinitions of various entanglement quantities like $T_E$ and $\mu_E$.

\subsection{Renormalisation of thermodynamic observables}
As we have seen that entanglement entropy of the strip system
gets corrected at higher orders in  perturbative calculation. It is
reasonable to expect that other thermodynamic variables  also receive
similar corrections at higher orders. We already saw that
the chemical potential $\mu_E$ indeed gets  corrected. Using second order
turning point expansion \eqn{kj4}, one can determine
\bea
\mu_E
&=& {1\over r_-}(2 {z_l^4\over z_\ast^4} -{z_l^4\over  z_0^4}) \br
&\simeq& {1\over r_-}\left( {2z_l^4\over \bar z_\ast^4} 
+{z_l^4\over  z_0^4}
[({4b_1\over b_0}-1) +{\bar z_\ast^4\over  z_0^4}(3b_2-5b_1^2)]\right) \br
&=& 
\mu_E^{Lif}
+ {1\over r_-}({4b_1}{\cal Z}-1){z_l^4\over  z_0^4} 
\eea
where ${\cal Z}=
1 +{\bar z_\ast^4\over  z_0^4}{(3b_2-5b_1^2)\over 4b_1}$. Thus 
the net difference in chemical potential due to excitations 
\bea
\bigtriangleup \mu_E^{(2)}
= {1\over r_-}({4b_1}{\cal Z}-1){z_l^4\over  z_0^4} 
\eea
has  second order terms also. 
It can be  seen that entropy expression \eqn{hj4nb} can be reexpressed as
a first law
\be
\bigtriangleup 
S_E^{(2)}=
 {1\over T_E}(\bigtriangleup E_R +{1\over 2} N_R 
\bigtriangleup \mu_E^{(2)})
\ee
The corresponding  entanglement temperature is given by
\be\label{temp1aw}
T_E=  
{64 z_l^2\over \pi (l_R)^3}
\simeq{64 z_l^2\over \pi l^3}{1\over (1- {1\over 35} { l^4\over(2z_0)^4})} 
\ee
which involves cubic power of renormalised length.
Similarly the `renormalised' energy and charge within subsystem 
 are also given in terms of $l_R$
\bea\label{hj5q}
&& \bigtriangleup 
 {E}_R
=  l_2 l_R {\cal E}, ~~~ N_R=l_2 l_R \rho,~~~
\eea
In  the above  `renormalised'  width of the strip is defined as
\be
l_R\equiv l ({ Q\over {\cal Z}})^{1\over 4}   \simeq
l \tilde Q
 \ee
where $\tilde Q =1-{1\over 105}{\bar z_\ast^4\over  z_0^4}$.
The new subsystem width $l_R$ includes terms only up to first order
${\bar z_\ast^4\over  z_0^4}$.
The result also suggests  $l_R< l$, which is consistent with the fact that
at the second order overall entanglement entropy decreases.   

In summary, in this universal approach
all  thermodynamic (extensive) quantities describing subsystem
are assumed to be dependent on 
the renormalised width through the volume factor.
It is also hypothesized that in terms of the corrected quantities
the entanglement first law should hold true at each order, like our approach
in \cite{ms2015}.
We also speculate
 that at the second order a new modular Hamiltonian  
can be inferred as
\be
\bigtriangleup 
S_E^{(2)}=
<\bigtriangleup 
H_E^{(2)}>.
\ee

\section{Summary}
We calculated the entanglement entropy of a strip like subsystem on 
the boundary of 
10-dimensional boosted black 3-brane solutions. These solutions 
when compactified along a lightcone coordinate  describe
excitations of 3-dimensional $a=3~ \theta=1$ hyperscaling Lifshitz  theory, 
at  fixed charge (momentum) density.
The theory has a natural scale $z_l$, determined by the charge density.
The area of the strip geometry for  
constant `lightcone time' is evaluated perturbatively up to second order.  
The  `finite' contribution of the $a=3~ \theta=1$
 Lifshitz ground state is found to be 
 $$S_{(0)}= 
{L^{3} \pi r_{-}l_2 \over 4 G_{5} z_l^2}\ln{l^2\over z_l^2}$$
 where an allowed range is
$l> 2z_l$. Due to the $z_l$ dependence,
this entropy is qualitatively different from the  2D CFT entropy 
which behaves as $\sim \ln (l/\epsilon)$ ($\epsilon$ 
being UV cutoff) \cite{RT}.  

The entanglement entropy of the excitations is however 
found to be proportional to $l^4$,  
whereas the  entanglement 
temperature falls off as ${1\over l^3}$. These results
are essentially along expected lines, indicating that the dynamical 
exponent of time for the hyperscaling Lifshitz background is three. Notably
these results are distinct  when compared
with the relativistic counterpart where the entanglement 
entropy of excitations instead grows as $l^2$, 
and temperature goes as ${1\over l}$, at first order \cite{JT}. 
A renormalisation 
of entanglement width is proposed   at second order 
when we try  to write down first law of  thermodynamics
$$\bigtriangleup 
S_E^{(2)}=
 {1\over T_E}(\bigtriangleup E_R +{1\over 2} N_R 
\bigtriangleup \mu_E^{(2)})
$$
This conclusion
falls  along the lines with the hypothesis invoked in an earlier 
work \cite{ms2015} for the relativistic case.       
The results in our paper can be generalised to
study higher dimensional  Lifshitz theories with varied 
dynamical exponents, for example
the cases listed in \cite{hs2012}, which  
follow from lightcone compactification of boosted black D$p$-branes
and black M$p$-branes.  

\vskip.5cm

\end{document}